\documentclass[conference]{IEEEtran}
\IEEEoverridecommandlockouts
\usepackage{cite}
\usepackage{amsmath,amssymb,amsfonts}
\usepackage{algorithmic}
\usepackage{graphicx}
\usepackage{textcomp}
\usepackage{xcolor}
\usepackage{balance}

\def\BibTeX{{\rm B\kern-.05em{\sc i\kern-.025em b}\kern-.08em 
    T\kern-.1667em\lower.7ex\hbox{E}\kern-.125emX}}
\begin{document}

\title{On the Impact of Classical and Quantum Communication Networks Upon Modular Quantum Computing Architecture System Performance\\
\thanks{This work was supported by the European Commission (QUADRATURE: 101099697). P.E. acknowledges support from an FPI-UPC grant funded by UPC and Banco Santander.}
}

\author{
    \IEEEauthorblockN{
        Pau Escofet\IEEEauthorrefmark{1},\
        Abhijit Das\IEEEauthorrefmark{1},\
        Sahar Ben Rached\IEEEauthorrefmark{1},\
        Santiago Rodrigo\IEEEauthorrefmark{1},\
        Jordi Domingo\IEEEauthorrefmark{1},\\
        Fabio Sebastiano\IEEEauthorrefmark{2},\
        Masoud Babaie\IEEEauthorrefmark{2},\
        Batuhan Keskin\IEEEauthorrefmark{3},\
        Edoardo Charbon\IEEEauthorrefmark{3},\
        Peter Haring Bolívar\IEEEauthorrefmark{4},\\
        Maurizio Palesi\IEEEauthorrefmark{5},\
        Elena Blokhina\IEEEauthorrefmark{6}\IEEEauthorrefmark{7}\
        Bogdan Staszewski\IEEEauthorrefmark{7},\
        Avishek Nag\IEEEauthorrefmark{7},\
        Artur Garcia-Sáez\IEEEauthorrefmark{8}\IEEEauthorrefmark{9},\\
        Sergi Abadal\IEEEauthorrefmark{1},\
        Eduard Alarcón\IEEEauthorrefmark{1},\ and
        Carmen G. Almudéver\IEEEauthorrefmark{10}
    }
    \IEEEauthorblockA{
        \textit{\IEEEauthorrefmark{1}Universitat Politècnica de Catalunya, Spain}
        \textit{\IEEEauthorrefmark{2}Delft University of Technology, The Netherlands}\\
        \textit{\IEEEauthorrefmark{3}Ecole Polytechnique Fédérale de Lausanne, Switzerland}
        \textit{\IEEEauthorrefmark{4}University of Siegen, Germany}
        \textit{\IEEEauthorrefmark{5}University of Catania, Italy}\\
        \textit{\IEEEauthorrefmark{6}Equal1 Labs, Ireland}
        \textit{\IEEEauthorrefmark{7}University College Dublin, Ireland}
        \textit{\IEEEauthorrefmark{8}Barcelona Supercomputing Center, Spain}\\
        \textit{\IEEEauthorrefmark{9}Qilimanjaro Quantum Tech, Spain}
        \textit{\IEEEauthorrefmark{10}Universitat Politècnica de València, Spain}\\
        pau.escofet@upc.edu
    }
    
}

\maketitle

\begin{abstract}
Modular architectures are a promising approach to scaling quantum computers beyond the limits of monolithic designs. However, non-local communications between different quantum processors might significantly impact overall system performance. In this work, we investigate the role of the network infrastructure in modular quantum computing architectures, focusing on coherence loss due to communication constraints. We analyze the impact of classical network latency on quantum teleportation and identify conditions under which it becomes a bottleneck. Additionally, we study different network topologies and assess how communication resources affect the number and parallelization of inter-core communications. Finally, we conduct a full-stack evaluation of the architecture under varying communication parameters, demonstrating how these factors influence the overall system performance. The results show that classical communication does not become a bottleneck for systems exceeding one million qubits, given current technology assumptions, even with modest clock frequencies and parallel wired interconnects. Additionally, increasing quantum communication resources generally shortens execution time, although it may introduce additional communication overhead. The optimal number of quantum links between QCores depends on both the algorithm being executed and the chosen inter-core topology. Our findings offer valuable guidance for designing modular architectures, enabling scalable quantum computing.
\end{abstract}

\begin{IEEEkeywords}
Quantum Computing Architecture, Quantum Communication, Distributed Quantum Computing
\end{IEEEkeywords}

\section{Introduction}
\label{sec:introduction}
Developing practical quantum computers capable of addressing real-world problems \cite{shor_polynomial_1997, Santagati2024} remains a complex task, with scalability being a key obstacle to overcome. Furthermore, ensuring the reliability of quantum computers requires implementing quantum error correction and fault-tolerant protocols \cite{10.1145/258533.258579} to mitigate noise and maintain computational accuracy, but these protocols come at the cost of significantly increasing the number of qubits required within the system. Multi-core (or modular) quantum computing architectures distribute the computational resources across multiple interconnected modules \cite{Bravyi_2022, jnane_multicore_2022, smith_scaling_2022, laracuente_modeling_2023}, and have emerged as a promising solution to scale quantum processors, paving the way for more powerful and resilient quantum computers.

Several key challenges must be addressed to transition from single-core (or monolithic) to multi-core quantum computing architectures. One crucial aspect is the development of an efficient communication infrastructure to enable the interaction of quantum states across quantum cores \cite{Escofet2023InterconnectFF}. Unlike classical communications, quantum communication introduces unique challenges due to the delicate nature of quantum states \cite{Wootters1982}. 

Entanglement-based quantum communication protocols require not only quantum links for distributing EPR pairs~\cite{PhysRev.47.777} (i.e. entangled states) among the quantum cores (or QCores) but also classical communication channels. For instance, when transferring quantum states using quantum teleportation \cite{gottesman_1999_demonstrating} (also known as \textit{TeleData}), classical communication is needed to transmit the 2-bit measurement outcomes. An alternative to quantum state transfer across cores is to perform a remote gate execution (also known as \textit{TeleGate}), where a two-qubit gate is executed between qubits located in different cores \cite{PhysRevA.62.052317}, also requiring entanglement generation and classical communications between the involved QCores.

While the literature on multi-core quantum computing has made significant strides in architecture design \cite{10.1145/3674151, laracuente_modeling_2023, caleffi2022distributedquantumcomputingsurvey}, a crucial aspect has been overlooked: the interplay of classical and quantum communications, and its impact on the performance of quantum systems. Since communication overheads can introduce latency and fidelity loss, optimizing the balance between computational resources and communication efficiency is essential for designing scalable modular quantum computers. A deep understanding of this interplay enables informed architectural choices. This is particularly important when allocating resources, as increasing communication capacity often comes at the expense of computational qubits, directly impacting the system's overall performance.

\begin{figure*}[t]
    \centering
    \includegraphics[width=\textwidth]{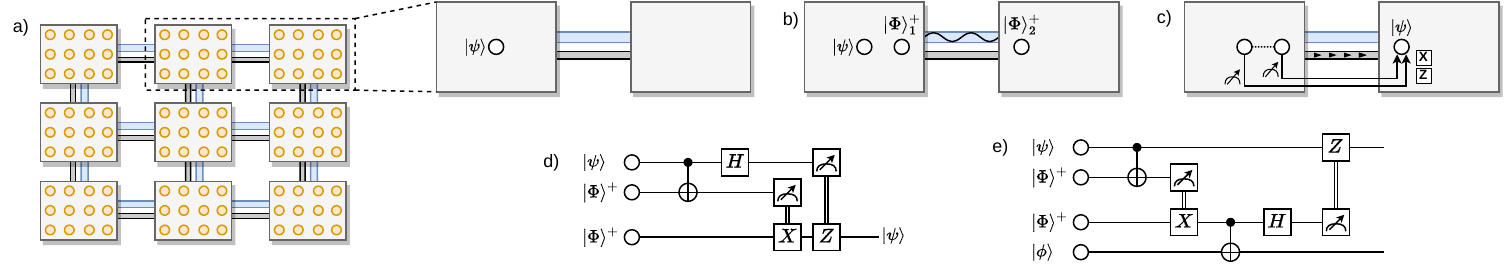}
    \caption{Multi-core quantum computing architecture based on the generation of EPR pairs. In this setup, QCores are connected through a quantum coherent link, which can generate an EPR pair (blue channel). Once these entangled states reside within the QCores, they are utilized (measured and thus consumed) for communication purposes, facilitating the transfer of quantum states across QCores. QCores are also connected through a classical network (black channel) to send the measurement outcomes (2 bits) required for the teleportation process. a) architecture overview with 9 QCores connected through a grid network. Quantum state $|\psi\rangle$ should be transmitted to the right core. b) and c) depict the use of each type of link (quantum and classical). In b) the quantum link generates and distributes an entangled pair. In c) teleportation is completed by measuring and transmitting the bits. d) depicts the whole teleportation circuit, and e) shows a remote CNOT executing between qubits $|\psi\rangle$ and $|\phi\rangle$, another protocol based on EPR pairs.}
    \label{fig:multi_core_architecture}
\end{figure*}

In this work, we explore how the classical and quantum networks impact the overall performance of modular quantum computing architectures, studying how different tradeoffs between computation and communication affect the system's fidelity. Specifically, our major contributions include:
\begin{itemize}
    \item An analysis of the latency introduced by the classical network for teleportation-based quantum transfers between QCores, identifying scenarios where it becomes the bottleneck.
    \item An exploration of various network topologies, evaluating how quantum communication resources impact the frequency and parallelization of inter-core qubit transfers.
    \item A full-stack evaluation of a proposed modular quantum computing architecture under different communication parameters and their impact on quantum fidelity.
\end{itemize}

The remainder of the paper is structured as follows. Section \ref{sec:background} briefly introduces modular quantum computing architectures and related quantum circuit mapping techniques. Section \ref{sec:full-stack} presents the modular architecture used in this work and explains the interplay between layers. Sections \ref{sec:methodology} and \ref{sec:results} describe the experiments conducted and summarize the findings. Lastly, Section \ref{sec:conclusions} presents the obtained conclusions and outlines potential paths for future research in this critical domain.

\section{Background and Related Work}
\label{sec:background}
\subsection{Multi-Core Quantum Computing Architectures}
\label{sec:multi-core_architectures}

Up to a thousand qubits have been integrated into a single processor in current quantum computers, which is a significant advancement in scalability \cite{atom2023kilo, chow_2021_ibm, gambetta_2023_the}. However, quantum computers need to be further enlarged to many thousands or even millions of qubits to be able to solve real-world problems \cite{preskill_quantum_2018}, even more if they need to incorporate error correction and fault-tolerant protocols \cite{nielsen_chuang_2010}.

Scaling up quantum computers to integrate that many qubits is a difficult endeavour. One fundamental challenge is maintaining low error rates while integrating precisely wired individual qubits with sophisticated classical control circuits \cite{NAP25196}. Furthermore, the pursuit of higher qubit counts needs to be achieved without increasing interference or crosstalk between qubits \cite{ding_systematic_2020}. As a result, increasing the number of qubits in a monolithic processor has a limit, and alternative solutions must be found.

A promising architectural approach involves scaling up quantum computation by interconnecting multiple quantum cores (QCores) \cite{Bravyi_2022, jnane_multicore_2022, smith_scaling_2022, laracuente_modeling_2023}, allowing the system to match the computational requirements of the algorithm while maintaining manageable unit sizes. Different levels of modularity are envisioned at increasing system complexity, all necessitating the implementation of communication methods. Specifically, for superconducting quantum processors, an initial proposal suggested connecting different chips via classical communication links \cite{Bravyi_2022}, aiming to execute large quantum algorithms demanding more qubits than available in a single core, in such case, circuit cutting and knitting techniques are required \cite{tang_2021_cutqc, piveteau2023circuit, carrera2024combining}. Subsequently, in the near future, short-range quantum links between neighbouring processors (\textit{i.e.}, chip-to-chip quantum coherent couplers) are anticipated to be introduced, enabling quantum communication via two-qubit gates across processors \cite{gold_entanglement_2021}.

As the field progresses, future advancements in multi-core quantum computing architectures anticipate a sophisticated integration of QCores facilitated by extended quantum-coherent communication links alongside classical channels. This progress will facilitate not only the coupling and entanglement of qubits distributed across various QCores within a single quantum computer \cite{gold_entanglement_2021, dijkema2025cavity, Qintranet}, but also the connectivity between separate quantum computing systems, transcending physical boundaries such as those between different refrigeration units \cite{magnard_microwave_2020}.

\begin{figure*}[t]
    \centering
    \includegraphics[width=\textwidth]{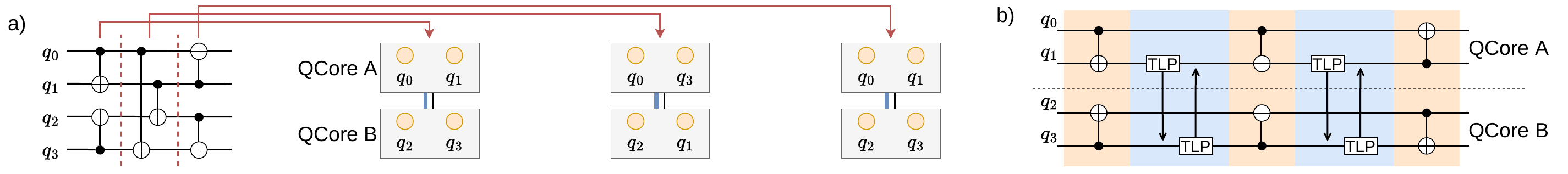}
    \caption{Quantum circuit mapping in multi-core architectures. a) Mapping a 4-qubit quantum circuit onto a 2-QCore quantum computing architecture. Within each timeslice, qubits are allocated to QCores, so every two-qubit interaction involves qubits placed within the same core. b) Compiled circuit with teleportations, each of them necessitating classical communications. Orange and blue regions represent computation and communication operations, respectively.}
    \label{fig:inter-core_mapping}
\end{figure*}

This work focuses on multi-core quantum computing architectures comprising multiple QCores capable of communicating thanks to the generation of EPR pairs ($|\Phi \rangle^+$) and their distribution to QCores through quantum links \cite{gold_entanglement_2021, magnard_microwave_2020, dijkema2025cavity}, as depicted in Figure \ref{fig:multi_core_architecture}. In this architecture, QCores are interconnected through quantum links (QLinks) that generate EPR pairs, which enable quantum state transmission between QCores via quantum teleportation \cite{gottesman_1999_demonstrating}. Figure \ref{fig:multi_core_architecture} d) shows the circuit necessary to perform quantum teleportation, including the transmission of measurement outcomes from the source QCore to the destination QCore. 

To ensure fast communication, it is imperative to establish the classical data transmission from QCore to QCore, bypassing the latency associated with routing the classical data through the host computer, the classical computer orchestrating the quantum operations.

Whether entangled states can be distributed between any pair of QCores or just a subset of them remains unanswered for now and will be addressed in future sections referred to as inter-core connectivity.

\subsection{Quantum Circuit Mapping for Distributed Quantum Computing}
\label{sec:background_circuit_mapping}

In single-core quantum processors, quantum circuit mapping techniques are used to efficiently assign each virtual qubit from the quantum circuit to a physical qubit of the quantum device and to ensure that every two-qubit gate from the circuit occurs between two adjacent physical qubits, complying with the qubit connectivity of the quantum processor (coupling map). Some examples of quantum circuit mappers for monolithic quantum computers can be found in \cite{li_2019_tackling, Qiskit, Amy_2020, lao_2022_timing, Sivarajah_2021, 10.1145/3569052.3578928, 10821308}.

Transitioning from single-core to multi-core architectures adds complexity to the mapping problem, introducing new challenges that depend on the interconnection topology of the QCores and the available communication primitives \cite{Escofet2023InterconnectFF}. As previously mentioned, this work assumes entanglement-based communication between the QCores, an EPR-distributed architectural model in which quantum states are transferred using classical and quantum channels. Throughout the circuit's execution, quantum states are relocated between QCores to guarantee that the corresponding qubits are positioned within the same QCore whenever a two-qubit gate needs to be executed. Therefore, the challenge lies in distributing qubits across QCores and routing them efficiently to minimize the number of non-local communications.

Figure \ref{fig:inter-core_mapping} a) illustrates the mapping of a circuit, consisting of four qubits and three timeslices, to a two-core architecture with two qubits per core. A timeslice refers to the disjoint set of quantum gates that can be executed simultaneously without overlapping qubits. Within each timeslice, each pair of interacting qubits must be allocated together in the same core, ensuring the feasibility of all two-qubit gates.

The movements across QCores will be performed using quantum teleportation between timeslices, yet the mapping process is not bound to teleportation itself, and any method of quantum state transfer can be applied after mapping the circuit into the architecture. The resulting augmented circuit is depicted in Figure \ref{fig:inter-core_mapping} b), where timeslices are divided into computation (orange) timeslices containing gates from the original quantum circuit and communication (blue) slices, where teleportation transfers will be added. 

Due to the still early development of modular architectures, just a few quantum circuit mapping techniques have been proposed. In \cite{baker_time-sliced_2020}, Baker et al. use a graph partitioning approach for solving the mapping problem in multi-core quantum computing architectures. Similarly, in \cite{bandic_mapping_2023}, Bandic et al. model the graph partitioning problem as a QUBO problem. Other heuristics models have been proposed, such as in \cite{Escofet_2023} and \cite{escofet_revisiting_2024}, where the Hungarian Qubit Assignment (HQA) is introduced and compared against the previously mentioned techniques. Other quantum circuit mapping algorithms for distributed quantum computing, such as \cite{ferrari_modular_2023, wu_autocomm_2022, andresmartinez2023distributing}, use remote gate execution apart from quantum teleportation.

\section{System Model}
\label{sec:full-stack}

\begin{figure}
    \centering
    \includegraphics[width=1\linewidth]{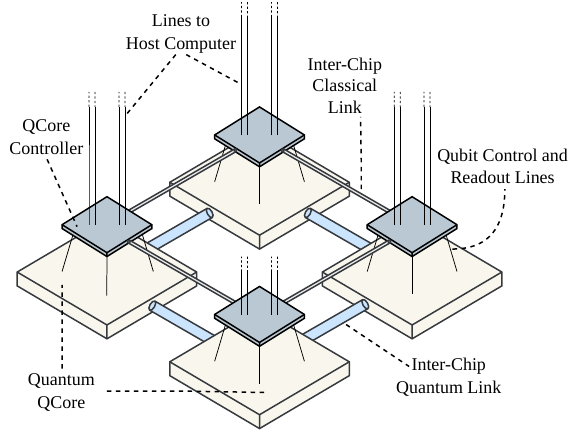}
    \caption{Schematic representation of the modular quantum computing architecture considered, showing four QCores connected in a ring topology. Each QCore is managed by a dedicated classical controller responsible for qubit control and readout. When a quantum state is transferred between two QCores, an entangled pair is first generated via the inter-chip quantum link (blue). The source QCore then performs preprocessing operations and measurements, with the resulting classical measurement outcomes transmitted to the destination QCore by its controller via the inter-chip classical link (gray). The destination QCore applies the necessary quantum corrections, completing the teleportation protocol.}
    \label{fig:modular_architecture}
\end{figure}

With the aim of analyzing the interplay between computation and communication in modular quantum computing architectures, we study a structured system model that captures the essential components of a modular architecture and their interactions. Figure \ref{fig:modular_architecture} depicts the considered architecture. 

The analyzed architecture distributes qubits across multiple QCores, each equipped with a dedicated classical controller responsible for local quantum operations and inter-core communications. A centralized host computer orchestrates circuit compilation and workload distribution at a higher level, ensuring efficient execution.

Inside each QCore, qubits are assigned specific roles: \emph{computational qubits} execute the quantum gates from the circuit, while \emph{communication qubits} facilitate inter-core interactions by storing halves of EPR pairs. These shared entangled pairs enable quantum teleportation, which we use to transfer quantum states between QCores as the circuit mapping dictates. To minimize latency, the classical communication required for teleportation occurs directly between QCore controllers. Specifically, each QCore controller transmits the necessary measurement outcomes to the destination QCore’s controller without routing through the host computer, thereby avoiding unnecessary delays.

QCores are connected via QLinks, which generate entanglement between qubits in neighbouring QCores. Various potential technologies for such links are under consideration, ranging from shared cryogenic environments \cite{gold_entanglement_2021} to mid-range links connecting processors housed in separate dilution refrigerators \cite{magnard_microwave_2020} and even long-distance optical links for distributed quantum computing \cite{wehner2018quantum}. However, since we focus on the logical organization of modular architectures, we abstract away the specifics of QLink implementation, only assuming that entanglement generation is possible. The following sections analyze the role of each architectural layer along with the assumptions made, covering the qubit plane and associated sources of decoherence, network infrastructure, circuit compilation, and system-wide performance trade-offs. 

\subsection{Qubit Coherence}
\label{sec:decoherence}
Quantum coherence is the ability of qubits to maintain superposition and entanglement. Decoherence disrupts these states, causing superpositions to collapse and destroying entanglement. Several factors contribute to qubit decoherence, including environmental noise and imperfections in qubit control mechanisms. These factors can perturb qubits, causing them to lose coherence rapidly. Two constants are typically measured to define quantum coherence over time: $T_1$ (thermal relaxation time) and $T_2$ (dephasing time). An exponential decay of the qubit coherence over time is assumed \cite{9201447, 10.1063/1.5089550}. We use these two terms to characterize how latency affects the coherence of quantum state \cite{escofet2025accurateefficientanalyticmodel} as:

\begin{equation}
    C(t) = e^{\frac{-t}{T_1}} \cdot (\frac{1}{2} e^{\frac{-t}{T_2}} + \frac{1}{2})
\label{eq:coherence}
\end{equation}

Considering how relaxation and dephasing times affect the final quantum state fidelity \cite{escofet2025accurateefficientanalyticmodel, t1_ibm, t2_ibm}.

Moreover, we assume each single- and two-qubit gate adds depolarizing noise \cite{nielsen_chuang_2010} to the quantum state, lowering its fidelity every time a gate is applied to the qubit.

The fidelity loss due to quantum gate operations in the circuit, referred to as operational fidelity ($F_{op}$), is estimated using the analytical model proposed in \cite{escofet2025accurateefficientanalyticmodel}. Instead of simulating the full quantum state, which would limit the architecture size to a few dozen qubits \cite{Boixo2018}, this model efficiently approximates fidelity by considering the cumulative effect of gate-induced noise, assumed to be a depolarizing error channel. By leveraging this approach, we can evaluate large-scale modular quantum architectures without the prohibitive computational cost of full-state simulations.

Both qubit decoherence over time ($C(t)$), computed as in Equation (\ref{eq:coherence}), and estimated operational errors ($F_{op}$) from \cite{escofet2025accurateefficientanalyticmodel} are combined into a single fidelity metric:

\begin{equation}
    F = C(t) \cdot F_{op}
\end{equation}

The execution's fidelity encapsulates all sources of errors (gate imperfections, measurement inaccuracies, and faulty entanglement generation) while also integrating communication-induced decoherence from latency. This provides a unified measure that reflects the overall performance of the full-stack modular architecture and serves as a single metric to be maximized.

\subsection{Network Infrastructure}
\label{sec:network_infrastructure}

\begin{figure*}
    \centering
    \includegraphics[width=1\linewidth]{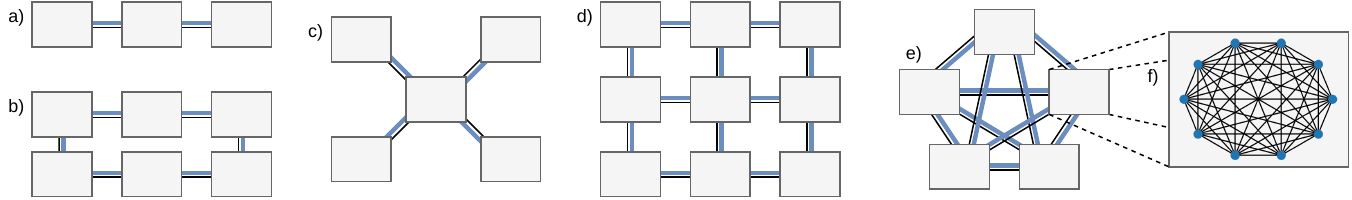}
    \caption{Used inter-core topologies. a) Line topology: QCores are connected in a 1D array manner. b) Ring topology: QCores are connected in a circular way. c) Star topology: All QCores are connected only to the central one. d) Grid topology: QCores are connected in a 2D array manner. e) All-to-all topology: All QCores are connected to all other QCores. f) Intra-Core topology. Inside each QCore, all qubits are directly connected to all other qubits in the same QCore.}
    \label{fig:inter-core_topologies}
\end{figure*}

In the design of multi-core quantum computing architectures, the choice of interconnection topology plays a pivotal role in determining the efficiency of quantum state transfer between QCores. Different topologies offer varying degrees of connectivity and ease of quantum state transfer, influencing the overall performance and scalability of the system.

The connection between two QCores indicates the possibility of generating and distributing an entangled state between them and transferring classical bits, thus performing quantum teleportation. If a quantum transfer is required between two QCores that are not directly connected, the quantum state will traverse through a path of intermediary QCores, with each hop necessitating one quantum teleportation.

In this work, we consider several inter-core topologies summarized in Figure \ref{fig:inter-core_topologies} \cite{dally2004principles, benini2002networks}. The \emph{Line topology} arranges the QCores in a 1D array, while the \emph{Ring topology} connects them in a circular manner. The \emph{Star topology} features a central QCore connected to all others, acting as a hub. The \emph{Grid topology} organizes QCores in a 2D array, and the \emph{All-to-all topology} ensures that every QCore is directly connected to all others. When evaluating the Grid topology for architectures necessitating $N$ QCores, these will be arranged in either a square ($\sqrt{N} \times \sqrt{N}$) or an almost-square ($\sqrt{N} \times (\sqrt{N} + 1)$) grid, ensuring a structured and balanced interconnection.

Since this study focuses on the impact of classical and quantum communications, the intra-core connectivity of each QCore (coupling map) is irrelevant to the work and is set to all-to-all connectivity. Thus, intra-core routing (\textit{i.e.}, adding local \texttt{SWAP} gates) is unnecessary. 

The network in the considered modular architecture is composed of a classical and a quantum network, both sharing the same connectivity. Once the topology is established, the routing algorithm defines the path quantum states follow, which can be (a) minimal or non-minimal and (b) deterministic, oblivious, or adaptive.

We employed minimal and deterministic routing algorithms across all inter-core topologies to ensure a fair evaluation, obtaining inspiration from classical networks \cite{dally2004principles, benini2002networks}. In the \emph{Line}, \emph{Star}, and \emph{All-to-All} topologies, a single path exists between any source and destination pair, making the routing inherently minimal and deterministic. For the \emph{Ring} topology, we used the shortest path, while for the \emph{Grid} topology, we applied the widely used XY-routing algorithm, where quantum states travel along the X dimension first, followed by the Y dimension, to reach their destination, \cite{enright2022chip}.

Since the communication infrastructure is shared, multiple QCores may compete for the same resource simultaneously. For instance, multiple qubits might request access to the same QLink to minimize latency in reaching their destinations. To resolve such conflicts, an arbitration policy is needed to select a winner. We prioritize transfers based on their remaining hop count to optimize overall system performance and minimize total system latency rather than individual packet delays. For example, if state $q_a$ is three hops away and state $q_b$ is five hops away from their respective destinations, $q_b$ is prioritized. By the time $q_b$ reaches its destination, $q_a$ will have also arrived, effectively hiding $q_a$'s latency within $q_b$'s. The same benefit does not apply if $q_a$ is prioritized. 

In the following sections, we evaluate each network's (classical and quantum) components and potential impact.

\subsubsection{Classical Communications Infrastructure}
\label{sec:classical_comms_infrastructure}
The classical communication infrastructure enables data transfer between QCores, typically using wired interconnects. To perform teleportation, we need to transfer the two measured bits in the source QCore, which will be used for the conditional execution of $X$ and $Z$ gates on the destination qubit, applying the necessary bit- and phase-corrections. Additionally, an identifier of the destination qubit ($\texttt{dst}_{id}$) must be sent to specify where these corrections should be applied. To enable acknowledgements and error correction methods in the classical domain, and since it is yet unclear how these systems will behave, we also include the source qubit identifier ($\texttt{src}_{id}$) in the data packet to be transmitted.

Each data transfer between two QCores requires two system clock cycles for routing and arbitration decisions. Packet transfer latency is further influenced by the number of parallel wired interconnects (or \emph{link width}) between QCores and the operating frequency of the communication infrastructure, which determines the system clock cycle. In this study, we adjusted the link width, defined as the number of parallel interconnects capable of transmitting one bit per cycle, and explored operating frequencies ranging from 10 MHz to 1 GHz, corresponding to system clock cycles between 100 ns and 1 ns. The impact of these variations on performance will be examined in the results section.

Considering the number of cycles ($\#\texttt{Cycles}$) and the clock frequency ($\texttt{freq}$), we compute the time it will take to transmit all the information for a quantum transfer to happen as:
\begin{equation}
    t = \#\texttt{Cycles}/\texttt{freq}
\end{equation}

Note that the number of cycles for the routing will depend on the link width and the packet size.

An identifier must be assigned to each physical qubit (regardless of the QCore where the qubit is located). Therefore, $\lceil \log_2 (\texttt{\#Qubits}) \rceil$ bits are needed for each identifier to target each possible physical qubit, making the packet size to transmit for each inter-core quantum state transfer as follows:
\begin{align}
    \texttt{packet size} &= \texttt{src}_{id} + \texttt{dst}_{id} + 2\\
        &= 2 \cdot \lceil \log_2 (\texttt{\#Qubits}) \rceil + 2
\end{align}



\subsubsection{Quantum Communications Infrastructure}
\label{sec:quantum_comms_infrastructure}
The quantum links (QLinks) in this work serve to generate and distribute entanglement between neighbouring QCores, enabling quantum state transfers via teleportation. Consequently, each QCore must allocate a portion of its qubits for communication purposes, reducing the number of qubits available for computation. The required communication qubits depend on the number of parallel QLinks between QCores.

\begin{figure}
    \centering
    \includegraphics[width=1\linewidth]{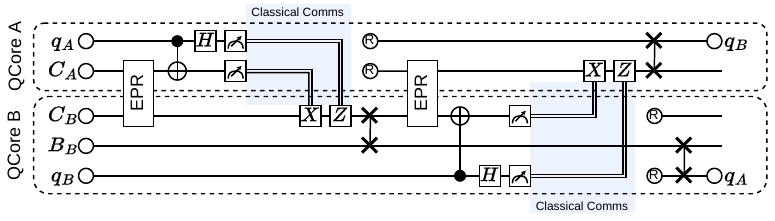}
    \caption{Two sequential teleportations using a shared QLink: the first transfers $q_A$ from QCore $A$ (top) to QCore $B$ (bottom), while the second transmits $q_B$ in the reverse direction. Three types of qubits are involved: computation qubits, which hold the quantum state being transferred ($q_{\{A,B\}}$); communication qubits, which store the entangled states ($C_{\{A,B\}}$); and buffer qubits, which temporarily hold the arriving quantum state ($B_{\{A,B\}}$). Notably, only two types of interactions are possible between QCores: entanglement generation and classical communication.}
    \label{fig:swap_tlp}
\end{figure}

Figure \ref{fig:swap_tlp} shows the circuit necessary to perform two sequential quantum teleportations by reusing communication resources. Note that a communication qubit ($C_{\{A,B\}}$) to store the entangled state and a buffer (or ancilla) qubit ($B_{\{A,B\}}$) to store the arriving state are used. Since the communication qubit is continuously used to generate new entanglement, it cannot be a storage medium for the arriving quantum state. Every QCore requires double the amount of QLinks in dedicated communication qubits, with one for generating entanglement and another serving as a buffer.

\begin{figure}
    \centering
    \includegraphics[width=0.8\linewidth]{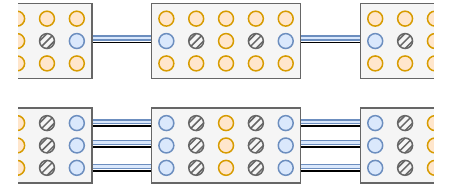}
    \caption{Illustration of a segment in a line-topology architecture, showing configurations with one (top) and three (bottom) parallel quantum links per connection. Each quantum link requires a dedicated communication qubit (blue) to hold entanglement and an ancillary qubit (grey) to serve as a buffer for incoming qubits. The remaining qubits within each QCore (orange) are allocated for computation.}
    \label{fig:comm_qubits}
\end{figure}

This constraint directly impacts scalability: a QCore with $Q$ total qubits and $l$ QLinks has $2l$ qubits reserved for communication, leaving $Q - 2l$ qubits for computation. If $2l > Q$, the QCore lacks the necessary space to accommodate the required qubits, causing the configuration to be unfeasible. Figure \ref{fig:comm_qubits} illustrates how this allocation changes with increasing numbers of QLinks, showing the computation qubits (orange) alongside communication (blue) and buffer (grey) qubits.

Ultimately, the number of QCores required to accommodate a computation of $n$ virtual qubits depends on the network topology and the number of parallel QLinks, highlighting the trade-off between communication capacity and computational resources.

\subsection{Quantum Circuit Mapping}
\label{sec:quantum_circuit_mapping}
As we move up the computation stack, we reach the compilation layer, which transforms high-level quantum algorithms into executable instructions tailored to the hardware constraints. The quantum circuit mapping process modifies a given quantum circuit with $n$ virtual qubits to make it executable on a target quantum system with $m \geq n$ physical qubits distributed across several QCores.

In our work, quantum circuits are mapped into multiple QCores, as depicted in Figure \ref{fig:inter-core_mapping}, using the HQA algorithm proposed in \cite{Escofet_2023}. This algorithm has been demonstrated to outperform other approaches~\cite{baker_time-sliced_2020, bandic_mapping_2023} in both the number of inter-core quantum state transfers and the execution time the algorithm takes, as shown in \cite{escofet_revisiting_2024}. This enables the system to increase the number of qubits and QCores while maintaining relatively low execution times. Moreover, the HQA algorithm allows us to define a restricted inter-core topology, dropping the assumption that QCores are connected in an all-to-all manner.

For constrained topologies, the HQA code has been modified such that the cost of moving a qubit $q$ placed in QCore $c_i$ to another QCore $c_j$ is set as the distance between them (\textit{i.e.}, $d(c_i, c_j)$).

In~\cite{Escofet_2023}, the cost function $\mathcal{C}_t (op_i, c_j)$ for assigning an operation $op_i$ into QCore $c_j$ involving qubits $q_A \in c_A$ and $q_B \in c_B$ is computed using the following equation:
\begin{equation}
    \mathcal{C}_t (op_i, c_j) = 
\begin{cases}
    \infty & \text{if } c_j \text{ is full}\\
    1 & \text{if } q_A \in c_j \text{ or } q_B \in c_j\\
    2 & \text{otherwise}
\end{cases}
\label{cost function with attraction}
\end{equation}

However, when dealing with restricted topologies, fixed costs for an inter-core transfer do not reflect the real cost of transferring one quantum state from one QCore to another. Hence, we calculate the cost using the distance, $d(c_i, c_j)$ between the QCores. Accordingly, Equation~\eqref{cost function with attraction} becomes:
\begin{equation}
    \mathcal{C}_t (op_i, c_j) = 
\begin{cases}
    \infty & \text{if } c_j \text{ is full}\\
    d(c_B, c_j) & \text{if } q_A \in c_j\\
    d(c_A, c_j) & \text{if } q_B \in c_j\\
    d(c_A, c_j) + d(c_B, c_j) & \text{otherwise}
\end{cases}
\label{cost function distance with attraction}
\end{equation}

Moreover, as explained in previous sections, in the proposed modular architecture, different QCores may host a different number of available qubits for computation, depending on the number of qubits per core, the number of communication qubits per link, and the inter-core topology. A clear example is the \emph{Star} topology with one QCore in the centre hosting fewer qubits devoted to computation due to its high inter-core connectivity compared to the other QCores on the architecture.

The HQA algorithm has been adapted so a different number of qubits can be assigned to each core, satisfying the available spaces and ensuring that the QCore capacity is never exceeded. Interested readers might refer to~\cite{Escofet_2023} and \cite{escofet_revisiting_2024} for more details on this multi-core quantum mapping algorithm.

It is important to note that quantum teleportation is not the only possible communication protocol for inter-core quantum links. For instance, remote gate execution \cite{PhysRevA.62.052317} enables the direct application of a two-qubit gate between distant qubits using the same resources as teleportation (one EPR pair and the transmission of classical information) without requiring qubit movement. Additionally, certain quantum links support direct state transfer \cite{magnard_microwave_2020}, allowing quantum states to be transmitted from one QCore to another without classical communication. However, this work focuses on teleportation, as entanglement generation is a common feature across most current inter-core quantum links, regardless of the underlying technology, and is a promising strategy for scaling up modular quantum networks \cite{wehner2018quantum}. Moreover, entanglement distillation \cite{bennett1996purification, pan2001entanglement} enables the enhancement of shared entangled states' fidelity through additional purification rounds, a feature not available in direct state transfer.

\subsection{Quantum Algorithm}
\label{sec:quantum_algs}
At the highest level of the computation stack, we reach the application layer, where quantum algorithms are executed to leverage the capabilities of the underlying hardware. To evaluate the performance of the modular quantum computing architecture described above, we employ four distinct quantum algorithms, each chosen to capture different computational and communication demands. The Quantum Fourier Transform (QFT) \cite{coppersmith2002approximate} features a high circuit depth and requires a high degree of qubit connectivity, making it particularly sensitive to inter-core communication overhead. The Quantum Volume (QVol) \cite{cross_2019_validating} benchmark consists of a randomly structured circuit designed to test quantum hardware's overall computational performance and error resilience. Lastly, the Greenberger-Horne-Zeilinger (GHZ) state preparation \cite{greenberger1989going} and the Cuccaro Adder \cite{cuccaro2004new} both exhibit a cascade-like structure with relatively short circuit depth.

These algorithms vary in both depth and structure, ranging from the highly regular interactions of GHZ to the entirely random connectivity of Quantum Volume.

All circuits have been obtained from the Qiskit library \cite{Qiskit}.

\section{Evaluation Methodology}
\label{sec:methodology}

\begin{table}
    \caption{Parameters and ranges used in the experiments.}
    \centering
    \begin{tabular}{c|c}
        \textbf{Parameter} & \textbf{Value}\\
        \hline
        \hline Single-qubit gate time & $7.9$ ns \cite{PRXQuantum.5.030353}\\
        \hline Two-qubit gate time & $30$ ns \cite{PhysRevLett.125.240503}\\
        \hline Measurement time & $40$ ns \cite{PhysRevApplied.17.044016}\\
        \hline EPR Generation time & $130$ ns \cite{gold_entanglement_2021} \\
        \hline
        \hline Single-qubit gate error & $7.42\cdot 10^{-5}$ \cite{li2023error}\\
        \hline Two-qubit gate error & $7\cdot 10^{-4}$ \cite{PhysRevLett.126.220502}\\
        \hline Measurement error & $1.67 \cdot 10^{-4}$ \cite{chen2023transmon, aasen2024readout}\\
        \hline EPR Generation error & $9 \cdot 10^{-3}$ \cite{gold_entanglement_2021}\\
        \hline
        \hline $T_1$ & $1.2$ ms \cite{PhysRevLett.130.267001}\\
        \hline $T_2$ & $1.16$ ms \cite{PhysRevLett.130.267001}\\
        \hline
        \hline Qubits per QCore & 64\\
        \hline Inter-Core Topology & Line, Ring, Star, Grid, All-to-all \cite{dally2004principles}\\
        \hline Comm. Qubits per Link & 1 -- 5\\
        \hline Classical Link Width & 1 -- 15\\
        \hline Clock Frequency & 10 MHz -- 1 GHz\\
        \hline Data Packet & $2\cdot \lceil \log_2(\texttt{\#qubits}) \rceil + 2 $\\
        \hline 
        \hline Logical Qubits & 256\\
        \hline Quantum Circuit & QFT, GHZ, Cuccaro, Quantum Volume
    \end{tabular}
    \label{tab:parameters}
\end{table}

\begin{figure*}
    \centering
    \includegraphics[width=1\linewidth]{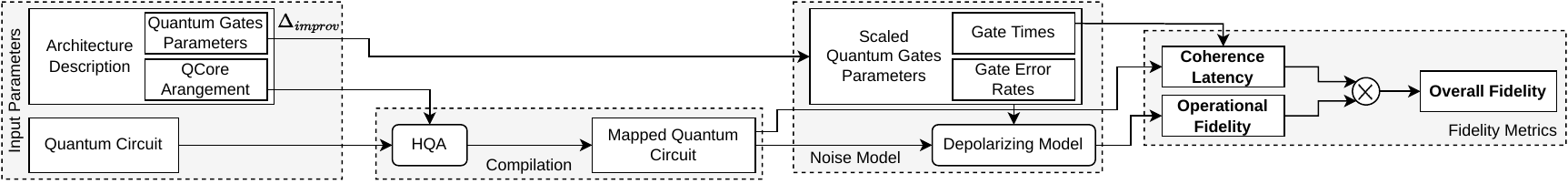}
    \caption{Proposed methodology's workflow. From input parameters (architecture description, quantum circuit, and $\Delta_{improv}$) to final fidelity metrics (Coherence Latency, Operational Fidelity, and Overall Fidelity), including compilation steps and noise model integration.}
    \label{fig:execution_flow}
\end{figure*}

To evaluate the feasibility and performance of modular quantum computing architectures, we analyze different inter-core topologies for a given quantum circuit with $n$ virtual qubits. Given a fixed QCore capacity (\textit{i.e.}, the maximum number of qubits per core) and a range of possible parallel QLinks between QCores, we determine the number of QCores required to allocate the circuit. As the number of QLinks increases, additional communication qubits are needed, which may sometimes exceed the available qubits per core, rendering specific configurations unfeasible.

We compile the quantum circuit using HQA \cite{escofet_revisiting_2024} for each feasible architecture to determine the number and timing of inter-core communications. Based on this, we schedule teleportations to minimize execution latency by leveraging the available QLinks. We estimate the final state fidelity for each circuit, which serves as our primary performance metric. Fidelity quantifies how closely the output state of the noisy execution matches the ideal target state, with higher values indicating better performance. Our objective is to maximize fidelity, balancing computational and communication errors while accounting for the impact of execution latency. By analyzing fidelity across different modular architectures, we gain insight into how inter-core communication, gate errors, and coherence loss affect the overall fidelity.

To ensure a comprehensive evaluation, we introduce the improvement factor $\Delta_{improv}$ for gate execution time and error rate. This prevents fidelity from converging to zero in large-scale simulations. $\Delta_{improv}$ is applied uniformly to both the duration of a quantum gate and its error rate, preserving the original fidelity loss ratio. That is, if the fidelity loss due to gate latency is half the loss due to the error rate for a single-qubit gate, the enhanced gate time and error rate under $\Delta_{improv}$ will maintain this ratio, leaving coherence times $T_{1,2}$ unchanged.
 
Fidelity is estimated using a depolarizing noise model, where each gate introduces a depolarizing error channel consistent with reported error rates. We employ the fidelity estimation method from \cite{escofet2025accurateefficientanalyticmodel}, incorporating errors from execution latency, as defined in Equation (\ref{eq:coherence}).

This methodology accounts for both latency-induced decoherence and direct fidelity losses from quantum operations. The proposed framework enables a detailed assessment of different architectures, providing insights into how various architectural components influence the final execution fidelity. Figure \ref{fig:execution_flow} provides an overview of the entire workflow.

Table \ref{tab:parameters} summarizes the architecture parameters used in the following experiments.

\begin{figure*}
    \centering
    \includegraphics[width=1\linewidth]{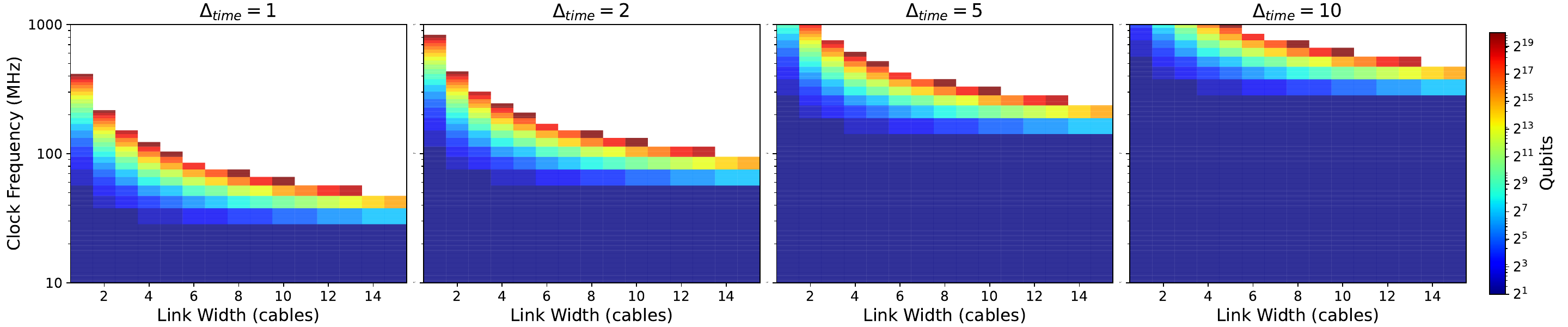}
    \caption{Architecture size (number of qubits in the system) at which classical communication becomes the dominant latency factor in a single quantum teleportation. The analysis sweeps transmission clock frequency, link width (number of wired interconnects), and $\Delta_{time}$ (improvement factor for gate, measurement, and EPR generation time) to identify the regime where classical delays outweigh quantum operation times.}
    \label{fig:classical_network_impact}
\end{figure*}

\section{Results}
\label{sec:results}
This section analyses how classical and quantum networks impact modular quantum computing performance. We begin by evaluating the classical network’s role in preserving quantum coherence and identifying scenarios where it becomes a bottleneck. Next, we assess the quantum network’s influence by analyzing the amount and parallelizability of non-local communications. Finally, we conduct a full-stack analysis to quantify the combined impact of these factors on system performance.

\subsection{Impact of Classical Network}
\label{sec:classical_comms_impact}
The classical network introduces latency through transmitting measurement outcomes, which dictates when X- and Z-flip corrections must be applied within the teleportation protocol. The transmission speed depends on clock frequency and link width, which determine the bit rate of communication. Faster transmissions reduce idling time and improve coherence preservation. However, it is still unclear how excessive wires or transmission rates may introduce thermal dissipation challenges in the cryogenic environment of quantum processors, potentially limiting scalability.

To identify when latency due to classical communication becomes a bottleneck, we compare classical communication times to the execution times of the other stages of quantum teleportation. Quantum teleportation can be decomposed into four key stages: (1) entanglement generation, where an EPR pair is created and distributed between the communicating QCores; (2) preprocessing, which includes the local quantum gates and measurements before transmitting classical information; (3) classical communication transmission, where the measurement outcomes are sent to the receiving QCore; and (4) postprocessing, which involves the correction operations and any necessary \texttt{SWAP} gates to complete the teleportation.

Figure \ref{fig:classical_network_impact} illustrates the system size (\textit{i.e.}, number of qubits) at which classical communication delays surpass the duration of the other teleportation stages, considering link widths from 1 to 15 wired interconnects and clock frequencies between 10 MHz and 1 GHz. Note that the classical packet size to be transmitted grows logarithmically with the number of qubits in the architecture. To account for potential technological advancements, we introduce $\Delta_{time}$, a parameter that models future improvements by reducing the execution time of gates, measurements, and entanglement generation per EPR pair by a factor of $\Delta_{time}$. As $\Delta_{time}$ increases, quantum operations become faster, increasing the relative impact of classical communication delays and highlighting when classical transmission becomes the dominant bottleneck in teleportation.

As shown in Figure \ref{fig:classical_network_impact}, for current gate execution times ($\Delta_{time} = 1$), a classical network operating at 100 MHz with 10 wires per link does not impose a bottleneck for system sizes exceeding one million qubits ($2^{20}$). However, as gate execution times are reduced, quantum operations accelerate while classical communication times remain unchanged. Consequently, the system size at which classical communication becomes the limiting factor decreases, highlighting its growing impact as quantum hardware improves.

\subsection{Impact of Quantum Network}
\label{sec:quantum_comms_impact}
Previous work on multi-core quantum mapping has primarily focused on minimizing the number of non-local communications \cite{escofet_revisiting_2024, baker_time-sliced_2020, bandic_mapping_2023, Escofet_2023}. Here, we extend this analysis by also considering the parallelization of these communications.

\begin{figure*}
    \centering
    \includegraphics[width=1\linewidth]{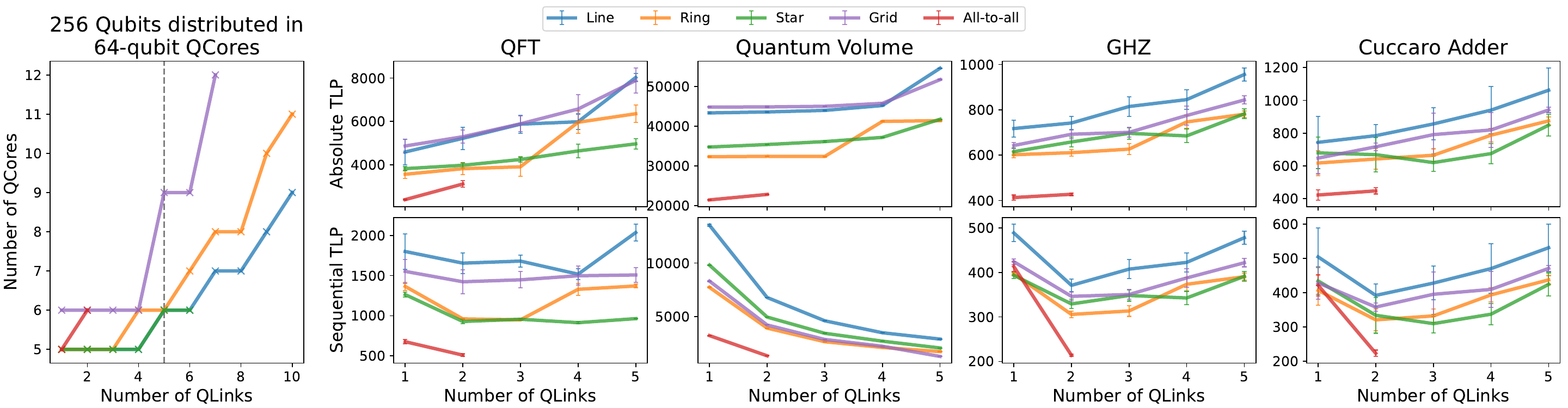}
    \caption{Resource requirements for executing the selected algorithms with 256 qubits in a modular architecture with 64-qubit QCores across different network topologies as the number of parallel QLinks increases (1--5). The figure reports the number of QCores (left), the absolute number of inter-QCore communications (upper row), and the number of sequential inter-QCore communications after parallelization (lower row). Results are averaged over five experiments, with error bars representing standard deviation.}
    \label{fig:quantum_network_impact}
\end{figure*}

\begin{figure*}
    \centering
    \includegraphics[width=1\linewidth]{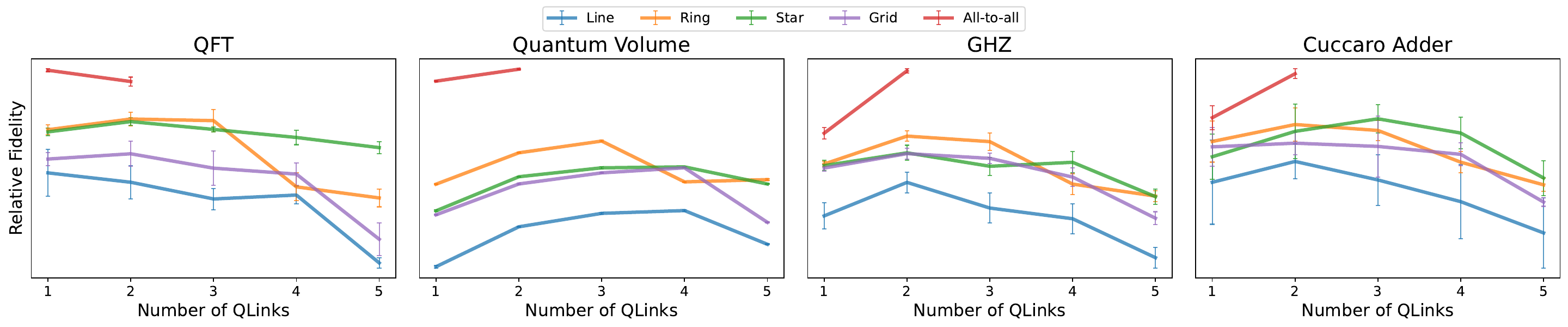}
    \caption{Relative fidelity estimates for four quantum algorithms (QFT, Quantum Volume, GHZ state preparation, and Cuccaro Adder) as a function of the number of parallel QLinks, which also determines the number of QCores. The improvement factor $\Delta_{improv}$ is set high enough to prevent fidelity from decaying to zero, enabling meaningful comparisons across architectures. Depolarizing noise is applied to gates according to reported error rates. Mean and standard deviation are calculated over five experiments.}
    \label{fig:alg_wise_fidelity}
\end{figure*}

Figure \ref{fig:quantum_network_impact} shows the impact of the quantum network when executing a 256-qubit circuit in a modular architecture with QCores containing up to 64 qubits. First, we determine feasible architectures for the given constraints. Next, we compile the circuit for each feasible configuration to quantify non-local communications. Finally, we schedule teleportations, identifying the non-local communications that can be executed in parallel when having more than one parallel QLink.

The results indicate that increasing the number of parallel QLinks generally reduces the number of teleportations contributing to latency ("\textit{Sequental TLP}") when the number of QCores remains fixed (\textit{e.g.}, from 1 to 4 parallel QLinks in a Line topology). However, if additional QCores are introduced (\textit{e.g.}, from 4 to 5 parallel QLinks in a Line topology), the total number of non-local communications rises, adding overhead. Moreover, even when the number of QCores remains unchanged, a higher number of QLinks can sometimes lead to increased latency for non-local communications. This occurs because the allocation of available qubit space within each QCore shifts, potentially requiring more non-local communications to execute a circuit. A full-stack analysis of the architecture is necessary to evaluate the trade-off between latency and the absolute number of non-local communications introduced by increasing the number of parallel QLinks.

\subsection{Full-Stack Analysis}
\label{sec:architecture_results}
Lastly, we analyze the entire modular quantum computing architecture, incorporating both classical and quantum communication constraints. Based on our previous findings, we set the classical link width to 10 and the clock frequency to 200 MHz, ensuring adequate bandwidth for a 256-qubit computation.

Previous experiments posed the question: \emph{Is it more beneficial to reduce the number of non-local communications or to lower the latency of those communications?}. Figure \ref{fig:alg_wise_fidelity} helps answer this question.

Figure \ref{fig:alg_wise_fidelity} presents the relative fidelity of the four benchmark quantum algorithms as the number of parallel QLinks increases. The considered architectural configurations correspond to those shown in the left plot of Figure \ref{fig:quantum_network_impact} (between 1 and 5 parallel QLinks), as the feasible architectures are determined by the number of virtual qubits (256) and qubits per QCore (64) rather than by the specific quantum circuit.

Instead of reporting absolute fidelities, which depend on the improvement factor ($\Delta_{improv}$) chosen, we present relative fidelities to facilitate a comparative analysis of different configurations. For these experiments, we set $\Delta_{improv}$ to a sufficiently high arbitrary value to prevent fidelity from decaying to zero, allowing us to focus on the relative performance of each configuration.

\begin{figure*}
    \centering
    \includegraphics[width=1\linewidth]{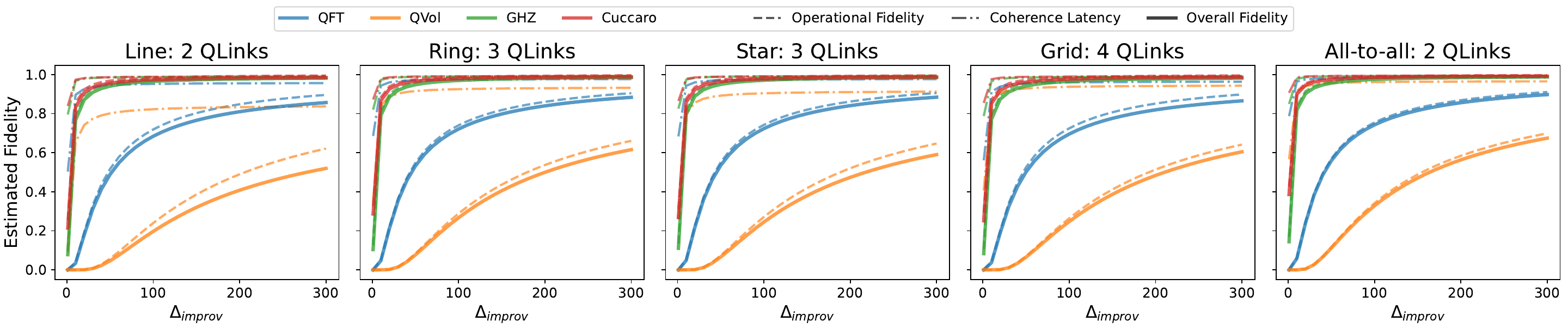}
    \caption{Estimated final fidelity for the selected benchmarks with 256-qubits distributed across 64-qubit QCores with two parallel QLinks for the Line and All-to-all topologies, three QLinks for the Ring and Star topologies, and four QLinks for the Grid topology. The selected QLinks for each topology is based on the previous analysis. The figure explores the fidelity estimation when increasing the improvement factor of gate times and error rates $\Delta_{improv}$. For each circuit topology pair, we report loss of fidelity due to gate errors (Operational Fidelity), loss of coherence due to computation and communication latency (Coherence Latency), and the combination of both (Overall Fidelity).}
    \label{fig:delta_exploration}
\end{figure*}

The results in Figure \ref{fig:alg_wise_fidelity} indicate that no single architecture consistently outperforms all others across all benchmarks. As expected, all-to-all connectivity among QCores yields the highest fidelity, as it minimizes the number of non-local communications required.

Take as an example the all-to-all topology. For the Quantum Volume, GHZ, and Cuccaro Adder algorithms, distributing the computation across six QCores connected in an all-to-all manner with two QLinks per connection yields the highest fidelity, even though one extra QCore is needed. However, for the QFT algorithm, the latency improvement does not compensate for the increased number of non-local communications introduced by additional QLinks. As a result, the highest fidelity is achieved when computation is distributed across five QCores connected in an all-to-all manner with just one QLink per connection.

Figure \ref{fig:delta_exploration} illustrates the effect of $\Delta_{improv}$ on the estimated fidelity for the selected benchmark circuits, where 256 qubits are distributed across 64-qubit QCores with different inter-core topologies. The overall fidelity is strongly influenced by the specific quantum algorithm, with GHZ and Cuccaro Adder circuits quickly approaching their maximum fidelity as $\Delta_{improv}$ increases.

A key observation is that coherence loss due to latency ("\textit{Coherence Latency}" in the figure) plateaus as $\Delta_{improv}$ increases, preventing fidelity from reaching 1 in certain cases. For instance, in the Line topology, the coherence due to latency for the Quantum Volume circuit stabilizes around 0.8. This showcases the impact of the classical network, which ultimately constrains execution time and imposes a fidelity ceiling. Regardless of improvements in quantum gate speed or reliability, achieving fidelities beyond this threshold needs improvements in the classical communication infrastructure.

Furthermore, coherence latency generally surpasses operational fidelity and rises rapidly with increasing $\Delta_{improv}$. This trend suggests that beyond a certain point, reducing gate execution time yields diminishing returns, while mitigating gate errors remains a crucial factor for scaling quantum computing architectures effectively.

\section{Conclusions and Future Work}
\label{sec:conclusions}
The scalability of quantum computers is one of the most pressing challenges in the field, as increasing the number of qubits while maintaining coherence and high fidelity is crucial for practical quantum computation. Modular architectures offer a promising pathway to scalability by distributing qubits across multiple QCores and utilizing quantum state transfer mechanisms (such as quantum teleportation) to enable communication between them. However, the efficiency of such architectures depends heavily on the underlying communication network (both classical and quantum).

In this work, we have proposed and evaluated a full-stack modular quantum computing architecture based on entanglement generation and quantum teleportation for quantum state transfer. We have examined the communication resources' impact on classical and quantum networks across five different topologies. Our analysis provides a detailed understanding of how network constraints influence the overall performance of modular architectures, shedding light on the trade-offs between communication overhead and computational capacity.

This study serves as a methodology for designing scalable modular architectures by quantifying how resources should be allocated between computation and communication. Since every architectural parameter (such as QCore size, number of QLinks, and network topology) affects system performance, understanding these trade-offs is essential for optimizing modular quantum computers. Our findings contribute to the broader effort of determining the most efficient ways to structure and scale modular quantum processors.

While the current analysis is constrained to relatively small qubit sizes due to computational limitations, the fidelity estimation pipeline developed in this work is fully scalable. Unlike classical simulations of quantum circuits, which become unfeasible for large systems, our methodology allows for analyzing larger platforms. Future work will extend this approach to analyze modular architectures at a significantly larger scale, further refining our understanding of communication-constrained quantum computing.

Additionally, integrating quantum error correction (QEC) and entanglement distillation into this framework represents a crucial next step. QEC can substantially reduce gate error rates by encoding logical qubits at the cost of increased execution time due to additional operations and syndrome extractions \cite{nielsen_chuang_2010, fowler2012surface}. Similarly, entanglement distillation can enhance the fidelity of shared EPR pairs used for quantum state transfer, though it introduces additional communication overhead and latency \cite{bennett1996purification, pan2001entanglement}. Our findings indicate that gate errors are the dominant source of decoherence in the studied architectures, suggesting that future research must carefully examine the trade-off between error rate reduction and execution time, both in the context of QEC (as a function of code distance) and entanglement distillation (balancing improved entanglement fidelity against increased latency).

This study highlights the importance of efficient communication in modular quantum architectures and provides a roadmap for their optimization. As quantum hardware advances, the insights gained here will help inform the design of scalable, high-performance quantum systems. With continued exploration of architectural trade-offs, modular quantum computing has the potential to overcome current scalability barriers and bring practical quantum advantage closer to reality.

\bibliographystyle{IEEEtran}
\balance
\bibliography{IEEEabrv, main}

\end{document}